\documentclass[prl,twocolumn,aps,amssymb,footinbib]{revtex4}
\usepackage{amssymb}
\usepackage{graphicx}
\usepackage{amsmath}
\usepackage{times}
\usepackage{color}
\usepackage{subfigure}
\usepackage{setspace}
\usepackage{bm}% bold math

\newcommand {\beq} {\begin{equation}}
\newcommand {\eeq} {\end{equation}}
\newcommand {\bqa} {\begin{eqnarray}}
\newcommand {\eqa} {\end{eqnarray}}

\newcommand {\up} {\ensuremath{\uparrow}}
\newcommand {\dn} {\ensuremath{\downarrow}}

\newcommand{\cd} {\ensuremath{c^\dagger}}
\newcommand{\ca} {\ensuremath{c^{\phantom \dagger}}}
\newcommand{\kk} {\ensuremath{{\bf k}}}

\newcommand {\qq} {\ensuremath{{\bf q}}}
\newcommand {\rr} {\ensuremath{{\bf r}}}

\begin{document}
\begin{abstract}
  We describe the dynamics of two component non-interacting ultracold
  Fermions which are initially in thermal equilibrium and undergo a
  rapid quench to either the repulsive or attractive side of a
  Feshbach resonance. The short time dynamics is dominated by the
  exponentially growing collective modes. We study the Stoner
  instability and formation of ferromagnetic textures on the repulsive
  side, and the pairing instability towards BCS or FFLO-like states
  (determined by the population imbalance) on the attractive side. In
  each case, we evaluate the growth rate of unstable modes and predict
  the typical lengthscale of textures to be formed.
\end{abstract}

\title{Non-equilibrium dynamics of interacting Fermi systems in quench
  experiments} 

\author{Mehrtash Babadi$^1$, David Pekker$^1$, Rajdeep Sensarma$^1$,
  Antoine Georges$^{2,3}$, Eugene Demler$^1$}

\affiliation{
  $^1$ Physics Department, Harvard University, Cambridge, Massachusetts 02138, USA\\
$^2$ Centre de Physique Theorique, Ecole Polytechnique, CNRS, 91128 
Palaiseau Cedex France\\
$^3$ College de France, 11 place Marcelin Berthelot, 75231 Paris Cedex 05, 
France 
}

\maketitle

Non-equilibrium dynamics of strongly interacting quantum many body
systems is a subject of great interest with diverse applications in
different fields, from condensed matter physics to high energy physics
to cosmology. However, compared to equilibrium physics, it lacks
deeper understanding from both theoretical and experimental
standpoints. Recent developments with ultracold
atoms~\cite{Bloch:Review} have opened a new window to study strongly
interacting model Hamiltonians. Although the focus has been mostly on
studying the equilibrium phases of these models, these systems also
present a unique opportunity to probe the intrinsic non-equilibrium
dynamics in the strongly correlated
regime~\cite{DefectProduction}. The low energy scales in the system
(typically $\sim$ kHz) and an almost ideal insulation from the
environment results in comparatively larger timescales of dynamics
than in the corresponding condensed matter systems and therefore, make
it easier to follow the intrinsic quantum evolution of the system in
real time~\cite{Strohmaier:Decay}.

The inherent tunability of interactions in these systems allows us to
study the dynamics of the system after a quench (rapid change) of the
Hamiltonian parameter. In fact, quench dynamics has already been studied 
in cold atom experiments that probe dynamics and pattern formation in spinor
BEC~\cite{SpinorBEC}, collapse and revival in optical
latices~\cite{Greiner2002}, and ``Bose novae'' in BEC's with
attractive interactions~\cite{Roberts2001}. Our primary motivation
for this study comes from recent experiments on cold two-component
Fermions in which interactions are tuned to the strong repulsive
regime during the course of the experiment~\cite{MIT:Stoner:Expt}.  It
was found that for sufficiently strong coupling, the atom loss
rate, the size of the atomic cloud and the distribution of the kinetic
energy of the atoms show signatures of a transition to a
ferromagnetic state. Originally, these experiments were interpreted as
probing the equilibrium Stoner transition~\cite{Stoner}.  On the other hand, the
time scales for tuning the interactions between small and large values
were limited by the formation of molecules on the BEC side of the
Feshbach resonance.  In this paper we show that the finite
experimental time scales are such that we can also analyze some
aspects of the experiments from the point of view of quench dynamics.

In this Letter, we focus on the initial dynamics of the collective
modes in Fermionic systems following a rapid quench.
Within linear response theory, we find that for quenches across a
critical interaction strength $U_c$, some collective modes become
dynamically unstable. We denote the growth rate corresponding to an
unstable mode with wavevector $q$ by $\Delta_q$. In general, the
lengthscale of the textures formed in the quench is bounded by the
smallest unstable wavelength $q_\text{cut}^{-1}$. Furthermore, if the
most unstable mode (i.e. for which $\Delta_q$ assumes its maximum) has
a finite momentum, $q_\text{max}$, then the order parameter will have
large spatial modulations at the corresponding wavelength
$q_\text{max}^{-1}$. We study two typical cases of dynamical
instability in Fermi systems starting with a non-interacting gas: (i)
quench to the regime of strong repulsive interactions which leads to
the Stoner instability and ferromagnetic textures; (ii) quench to the
regime of weak attractive interactions and the BCS pairing
instability. The order parameter in the former case (magnetization)
is conserved, while in the latter case (wavefunction of the coherent pairs), it is not conserved.

Our main results are: (a) for the Stoner instability in 3D, the
wave-vector of the most unstable mode scales as $q_\text{max}\sim
u^{1/2}$ while its growth rate $\Delta_\text{max} \sim u^{3/2}$, where
$u= U_f/U_c-1$ and $U_f$ the final coupling after the quench. (b) In
2D, at $T=0$, the most unstable mode has growth rate $\sim u$ and
occurs at $\sqrt{2}k_F$, where $k_F$ is the Fermi wave-vector of the
gas. However, the system is very sensitive to thermal fluctuations and
at finite temperatures, $q_\text{max}$ scales with $u^{1/2}T^{1/2}\exp
(\epsilon_F/2T)$ while the growth rate $\sim u~q_\text{max}$. This
scaling reverts back to its $T=0$ form once $q_\text{max} \sim
\sqrt{2}k_F$. (c) In the BCS case (in 3D), we find that for small
initial spin imbalances, the fastest growing instability is towards
formation of coherent Cooper pairs with center of mass momentum
$\qq=0$. Beyond a critical polarization, the fastest growing modes
correspond to pairings at a finite wavevector, i.e. towards FFLO-like
states~\cite{FFLO, FFLOdynamics}. Beyond a $2^{nd}$ critical
polarization, all modes become stable. We note that the change in the
character of dynamics does not necessarily correspond to crossing a
thermodynamic phase transition during the quench; e.g. although there
is no ferromagnetic transition in 2D at finite temperatures, quench
dynamics still involves unstable modes which exhibit exponential
growth.

\begin{figure}
\includegraphics[width=8.5cm]{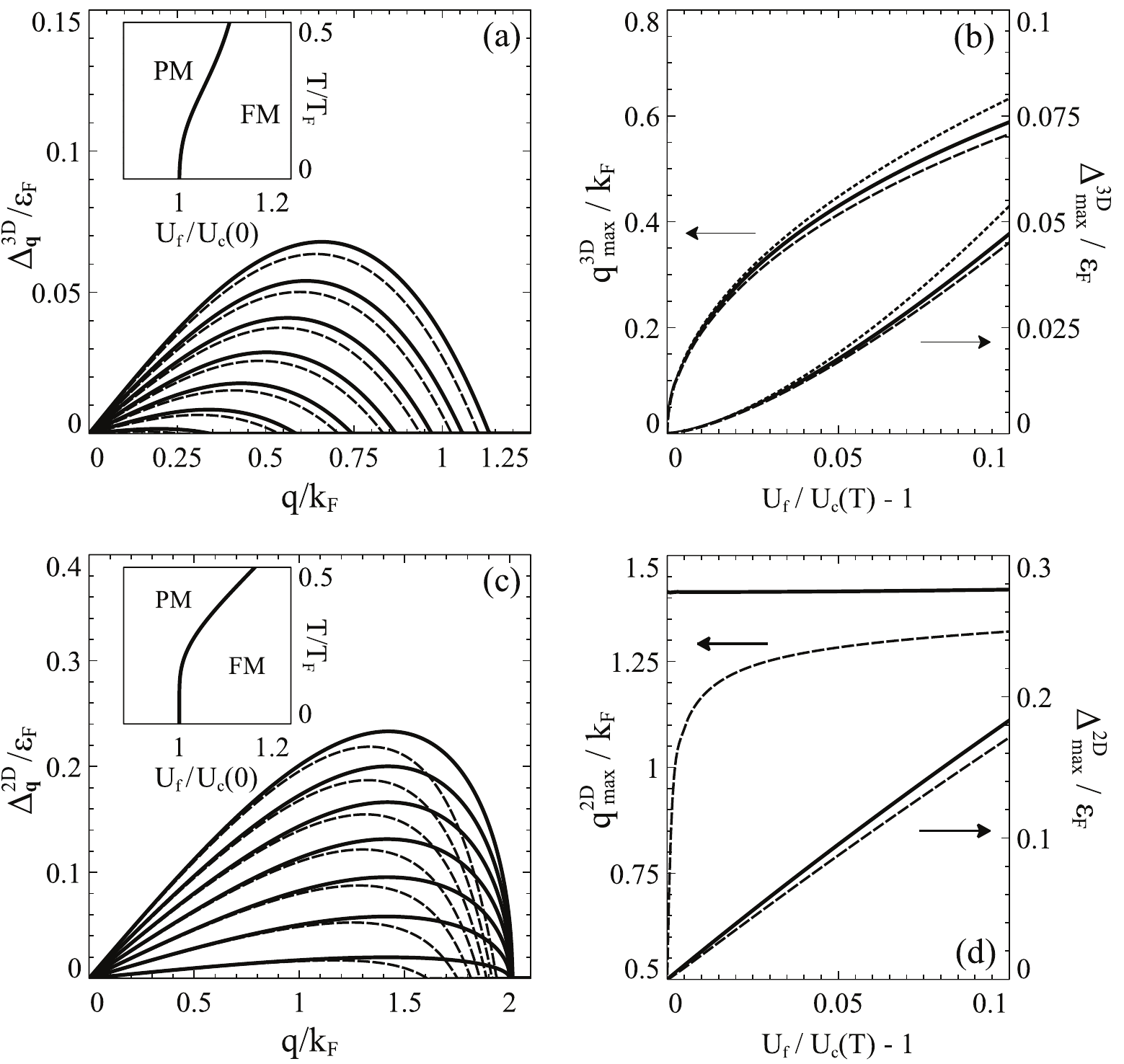}
\caption{Properties of growing collective modes in the Stoner
  instability in 3D (top row) and 2D (bottom row). (a) and (c): Growth
  rate $\Delta_q$ as a function of wavevector $q$ for $T=0$ (thick
  lines) and $T=0.1\,T_F$ (dashed lines) and several values of the dimensionless coupling $u=U_f/U_c(T)-1=0.01$ (bottom pair of curves),
  $0.03, 0.05, \ldots, 0.13$ (top pair of curves). The insets show the
  dynamics phase diagram as a function of final interaction strength
  $U_f/U_c(0)$ and reduced temperature $T/T_F$. ``PM'' corresponds to
  the case where all collective are damped, while ``FM'' corresponds
  to the existence of an unstable collective mode.  (b) and (d): The
  most unstable wavevector $q_\text{max}$ and the corresponding growth
  rate $\Delta_\text{max}$ vs. the dimensionless coupling
  $u$ for $T=0$ (solid lines, both 2D and 3D), $T=0.2 \,
  T_F$ (dotted lines, 3D), $T=0.1\,T_F$ (dotted lines, 2D). The dashed
  lines in (b) are the approximate analytical solutions. }
\label{fig:stoner}
\vspace{-0.5cm}
\end{figure} 

{\it Formalism}-- We consider a system of interacting Fermions
described by the Hamiltonian:
\begin{align}
H=\sum_{\kk,\sigma} \xi_{\kk \sigma} \cd_{\kk\sigma} \ca_{\kk\sigma}  + 
U(t) \int d^d\rr \, \cd_{\rr \up} \cd_{\rr \dn} \ca_{\rr \dn} \ca_{\rr \up},
\label{eq:H}
\end{align}
where $\cd_\sigma (\ca_\sigma)$ are the Fermion creation
(annihilation) operators with spin $\sigma$, $\xi_{\kk
  \sigma}=k^2/2m-\mu_\sigma$, $m$ is the mass of the Fermions,
$\mu_\sigma$ are the chemical potentials, and $U(t)$ is the time
dependent coupling. $U>0$ corresponds to the repulsive case (Stoner
instability), while $U<0$ corresponds to the attractive case (BCS
pairing instability). We focus on the instantaneous quench limit, in
which the coupling $U$ changes from a negligible initial value $U_i$
to a final value $U_f$ at time $t=0$.  $U_f$ is chosen to lie beyond
the critical interaction strength $U_c(T)$ for the transition in
question. The instantaneous quench approximation is valid if the time
of ramp is much shorter than the growth rate of the instability (will
be discussed later). In rapid quenches the distribution of
fluctuations present in the system does not change during the quench,
and is given by the equilibrium distribution at the initial value of
interaction $U_i$.  We are interested in the dynamics of this initial
distribution after the quench. This dynamics is dominated by unstable
collective modes which grow exponentially in time and lead to
formation of textures. The equation of motion for the collective
coordinate $\hat{\Phi}_\qq$ is given by:
\begin{align}
i \partial_t \hat{\Phi}_\qq(t)=[\hat{\Phi}_\qq(t),H],
\end{align}
where $\hat{\Phi}_\qq=S^z_\qq=\frac{1}{2}\sum_{\kk\sigma} \sigma
c^\dagger_{\kk+\qq,\sigma} c^{\phantom{\dagger}}_{\kk,\sigma}$ for the
Stoner instability, while for the pairing instability,
$\hat{\Phi}_\qq(t)=\sum_\kk c^\dagger_{\kk+\qq,\uparrow}
c^\dagger_{-\kk,\downarrow} $. To probe the dynamics, we couple an
external field $B_\qq(t)$ to the collective coordinate and look at its
response to this perturbation (for Stoner, the external field
is just the magnetic field while for BCS, it is a
fictitious pairing field). Within the framework of RPA,
we obtain:
\begin{align}
\Phi_\qq(t)=\int \chi^0_\qq(t-t') \left[ U(t') \Phi_\qq(t')+B_\qq(t')\right],
\label{Eq:prop0}
\end{align}
where $\Phi_\qq(t)=\langle \hat{\Phi}_\qq(t) \rangle$ and
$\chi^0_\qq(t-t')$ is the retarded bare susceptibility in the
initial (unordered) equilibrium state. This assumes that the Fermion
distribution functions do not change during the short time
dynamics after a rapid quench. Eq.~(\ref{Eq:prop0}) may be interpreted
as the propagation of the total field (composed of the
internal mean field $U(t') \Phi_\qq(t')$ and the external field
$B_\qq(t')$) forward in time by $\chi^0_\qq(t-t')$.
If the external field $B_\qq(t')$, as well as $\Phi_\qq(t')$ vanish for
$t' \leq 0$~\cite{init_footnote}, Eq.~(\ref{Eq:prop0}) can be
explicitly solved for $\Phi_\qq(t)$:
\begin{align}
\Phi_\qq(t)=\int dt' \, \chi_\qq(t-t';U_f)  B_\qq(t'). 
\end{align}
where $\chi_\qq(t-t';U_f)$ is the inverse Fourier transform (with the
integration contour shifted above all poles in the complex plane) of
\begin{align}
\chi_\qq(\omega;U_f)=\frac{\chi^0_\qq(\omega)}{1-U_f \chi^0_\qq(\omega)}.
\label{Eq:RPA}
\end{align}
This useful result relates the short time dynamics of the collective
mode after a quench to an RPA-like susceptibility in which the bare
response functions are evaluated in the {\em initial state}, but with
a coupling corresponding to the {\em final state}. When
$\chi_\qq(\omega_\qq;U_f)$ has poles at $\omega_\qq=\Omega_\qq+ i
\Delta_\qq$ in the upper half of the complex plane, then fluctuations
that occur after the quench will grow exponentially in time. In the
next sections, we make these ideas explicit by studying the specific
cases of the Stoner and the BCS instability.

One of the common concerns about the existence of the Stoner
instability in equilibrium systems is the so called Kanamori
argument~\cite{Kanamori}, which says that the effective $U$ which
enters the Stoner criterion gets reduced by screening. We expect that
the time scale associated with the osscillations of the screening
potential is $\sim \epsilon_F^{-1}$ which is much faster than
$\Delta_\qq^{-1}$. We proceed by assuming that there is a phase
transition, so, the $U_f$ that enters Eq.(~\ref{Eq:RPA}) should be
considered to be the renormalized $U_f$.

{\it Stoner instability}-- For a quench of the interaction across the
Stoner instability, the magnetization propagator is given by
Eq.~(\ref{Eq:RPA}), with
\begin{equation}
\chi^0_\qq(\omega) = \int \frac{d^{d}\mathbf{k}}{(2\pi)^d}
\frac{n^F(\xi_{\kk+\qq})-n^F(\xi_\kk)}{\omega-(\xi_{\kk+\qq}-\xi_\kk)+i0^+},
\label{Eq:chi0Stoner}
\end{equation}
where $\chi^0_\qq(\omega)$ is the bare spin susceptibility per spin
of a free Fermi gas, and $n^F$ is the Fermi distribution function. The
poles of the corresponding propagator are then found from:
\beq
1-U_f\chi^0_\qq(\Omega_\qq+i\Delta_\qq)=0.
\label{pole:eqn}
\eeq
A line of complex poles with a positive imaginary part $\Delta_\qq$,
appears when $U_f$ exceeds $U_c(T)$ which corresponds to the Stoner
instability (see insets of Fig.~\ref{fig:stoner}a and c).
$\Delta_{\qq}$ grows linearly for small $q$, reaches a maximum value,
$\Delta_{\mathrm{max}}$ at a wave-vector $q_{\mathrm{max}}$, and
decreases to zero at $q_{\mathrm{cut}}$. Modes with
$q>q_{\mathrm{cut}}$ are stable. 

We begin by considering the 3D case. At $T=0$, the integrals in
Eq.~(\ref{Eq:chi0Stoner}) may be evaluated explicitly, yielding the
familiar Linhardt function, analytically continued to the upper
half-plane. We obtain the approximate expression for $\Delta_\qq$ at
$T=0$: $ \Delta_\qq^{\mathrm{3D}}=(2k_F/\pi m) u q - (1/6 \pi m k_F)
q^3 + \mathcal{O}(q^5) $.  Fig.~\ref{fig:stoner}a shows $\Delta_\qq$
as a function of $q$, for $T=0$ and $T=0.1~T_\text{F}$ and several
different values of $u=U_f/U_c(T)-1$. In all cases $\Delta_{\qq=0}=0$
as required by the conservation of total magnetization. The values of
$\Delta_{\mathrm{max}}$ and $q_{\mathrm{max}}$ are plotted as a
function of $u$ in Fig.~\ref{fig:stoner}b. The analytic expression for
$\Delta_\qq$ gives $ q_{\mathrm{max}}^{\mathrm{3D}} \simeq 2 k_F
u^{1/2}$ and $\Delta_{\mathrm{max}}^{\mathrm{3D}} \simeq (16/3\pi)
\epsilon_F u^{3/2}$. These approximate locations of
$q_{\mathrm{max}}$ and $\Delta_{\mathrm{max}}$ appear in
Fig.~\ref{fig:stoner}b with dashed lines, and show a good agreement
with the exact numerical solution of Eq.~(\ref{pole:eqn}) at $T=0$.
The scaling laws correspond to the critical exponents obtained for the
equilibrium paramagnetic-ferromagnetic transition with $\nu=1/2$ and
$z=3$~\cite{Vojta}. The temperature dependence of the above quantities
have also been studied numerically and are plotted in
Fig.~\ref{fig:stoner}a and \ref{fig:stoner}b as dotted lines. We note
that the scaling laws remain unchanged at finite temperatures.

Next, we consider the same instability in 2D, which is the lower
critical dimension. In 2D, there is a quantum phase transition to the
ferromagnetic state at $T=0$, but there is no thermodynamic phase
transition at finite temperatures. We find, however, that even at
nonzero temperatures, there still is a critical interaction strength,
$U_c(T)$, such that for $U_f>U_c(T)$, there are exponentially growing
collective modes in the short time dynamics (see the inset of
Fig~\ref{fig:stoner}c).  At $T=0$,
Eqs.~(\ref{Eq:chi0Stoner}) and (\ref{pole:eqn}) yield
$\Delta_\qq^{\text{2D}} = \epsilon_F (q/k_F) [u/(1+u)]\sqrt{3-(q/k_F)^2 + 2u
  +1/(2u+1)}$, from which we get $\Delta_{\mathrm{max}}^{\mathrm{2D}}
= 2u\epsilon_F/(2u+1) \simeq 2 \epsilon_\text{F} u$ and
$q_{\mathrm{max}}^{\mathrm{2D}} = k_F \sqrt{u+\left[3
    +1/(2u+1)\right]/2} \simeq \sqrt{2} k_F$. In
Fig.~\ref{fig:stoner}c and~\ref{fig:stoner}d we plot the same
quantities as in Fig.~\ref{fig:stoner}a and~\ref{fig:stoner}b, but for
the 2D case. The major difference in 2D is that, at finite
temperatures, $\qq_{\mathrm{max}}^{\mathrm{2D}}$ starts at $\qq=0$ for
$u=0$ and rapidly grows to $\sqrt{2}k_F$ as one quenches deeper into
the unstable side. From a low temperature expansion, we find that at
$T\neq 0$, $\qq_{\mathrm{max}}^{\mathrm{2D}}\sim u^{1/2}(T)^{1/2}
\exp(\epsilon_F/2T)$ while $\Delta_{\mathrm{max}} \sim
u^{3/2}(T\epsilon_F)^{1/2}\exp(\epsilon_F/2T)$ for small $u$. These
scaling solutions go over to the $T=0$ scaling forms when $u\sim
(\epsilon_F/T) \exp(-\epsilon_F/T)$, which is an exponentially small region
at small $T$.

\begin{figure}
\includegraphics[width=8.5cm, bb=0 0 441 214]{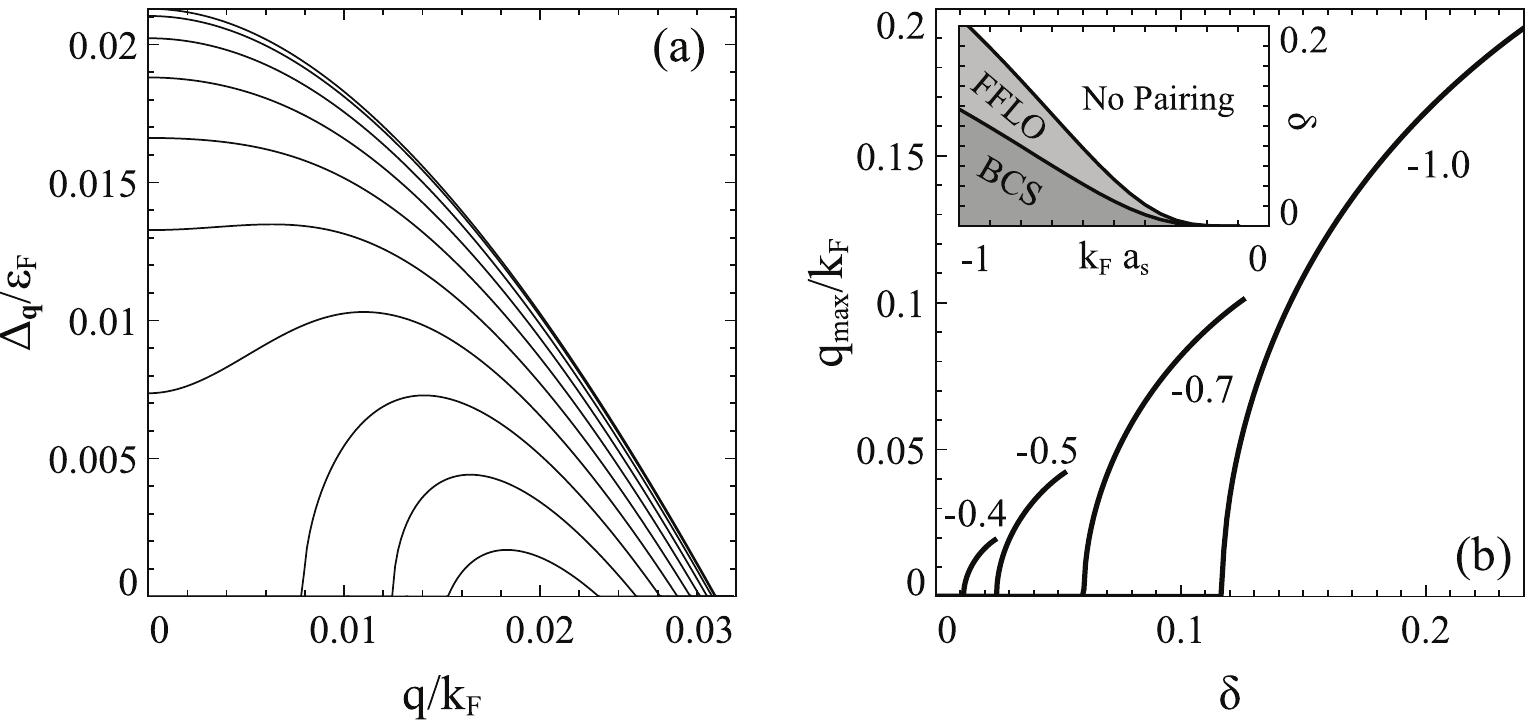}
\caption{Unstable collective modes in the BCS instability in 3D. (a)
  Growth rate of modes, $\Delta_q$, as a function of wavevector $q$
  for various values of the population imbalance $\delta=0$ (top
  curve), 0.0025, \ldots, $0.0225$ (bottom curve); ($k_F a_s = -0.4$).
  (b) The most unstable wavevector $q_\text{max}$ as a function of
  $\delta$ for various values of the $k_F a_s$ (indicated). As
  $\delta$ increases, $q_\text{max}$ goes through a continuous
  transition; the curves terminate at large $\delta$, above which all
  pairing modes are stable. Inset: the phase diagram in the $k_F
  a_s$-$\delta$ plane, showing the character of the most unstable
  mode.}
\label{fig:BCS}
\end{figure}

{\it Pairing instability}-- In this section we consider the
instability of a 3D Fermi gas with attractive interactions at
$T=0$. The case with no spin imbalance has been studied in the
literature before~\cite{AGD}. Here, we study the general case with
spin imbalance~\cite{Zwierlein,Polini}. We start by noting that the bare BCS
susceptibility has a ultraviolet divergence which is regularized by
expressing the microscopic interaction strength, $U_f$, in terms of
the s-wave scattering length~\cite{Randeria93}. Following this
regularization procedure, we obtain an expression for the poles of the
propagator of the pairing fluctuations [the equivalent of
Eq. (\ref{pole:eqn})]:
\begin{equation}
\frac{m}{4 \pi a_s}= \int \frac{d^{d}\mathbf{k}}{(2\pi)^d}
\left(\frac{1-n^F(\xi_{\kk,\up}) - n^F(\xi_{-\kk+\qq,\dn})}{\omega_\qq - \xi_{\mathbf{k}\up} - \xi_{\mathbf{-k+q}\dn}+i0^+}+\frac{m}{\kk^2}\right),
\label{Eq:poleBCS} 
\end{equation}
where $\omega_\qq=\Omega_\qq+i\Delta_\qq$, and $a_s < 0$ is the final
s-wave scattering length. In Fig.~\ref{fig:BCS}a, we plot $\Delta_\qq$
as a function of $q$, for several values of spin imbalance
$\delta=(n_\up-n_\dn)/(n_\up+n_\dn)$ and a final s-wave scattering
length $k_F a_s = -0.4$.  For the spin-balanced case we recover the
approximate solution $\Delta_\qq \approx \Delta_0 - 2q^2/3
k_F^2\Delta_0-\mathcal{O}(q^4)$, where $\Delta_0 = 8\epsilon_F
\exp(\pi/2 a_s-2)$, from which we infer $q_{\text{cut}} = (\sqrt{3/2})
\Delta_0 k_F$. We note that $\Delta_0 = \Delta_{\text{BCS}}$, i.e. the
growth rate of the BCS type pairings is equal to the BCS gap at
equilibrium~\cite{AGD}. Thus, for the spin balanced case the growth
rate decreases monotonically with $\qq$ and the strongest instability
is towards pairing with zero center of mass momentum.  The same
picture holds at small but finite population
imbalances~\cite{Tomadin2008}. However, as $\delta$ crosses a critical
value, the maximum of the growth rate continuously shifts to a nonzero
$q$. Thus, for large enough $\delta$, the short time dynamics is
dominated by finite-momentum pairings. This is illustrated in
Fig.~\ref{fig:BCS}b where we plot the wave-vector of the most unstable
state, $q_\text{max}$, as a function of spin imbalance $\delta$ for
several values of $k_F a_s$. As the imbalance is increased further,
the zero momentum pairing instability disappears completely
(corresponding to Clogston-Chandrasekhar
limit~\cite{Clogston:Chandra}). Finally, all pairing instabilities
disappear at larger population imbalances, which corresponds to the
sudden end of the curves in Fig.~\ref{fig:BCS}b at large $\delta$.

In the inset of Fig.~\ref{fig:BCS}, we have plotted the phase diagram
of the system as determined by the nature of the dominant unstable
mode in the coupling ($k_F a_s$) and the spin imbalance, ($\delta$)
plane. Note that FFLO is used in a loose sense of finite momentum
pairing while BCS corresponds to zero momentum pairing. This diagram
resembles the mean-field {\em equilibrium} phase diagram~\cite{FFLO},
with the significant difference being a larger region dominated by
finite-momentum pairings. We finally note that instead of directly
probing the pairing order parameter, one can look for spin
correlations associated with finite wave-vector pairings through
interactions in the Cooper channel, which are expected to appear for
quenches with sufficiently large spin imbalances.

{\it Comment regarding lattice}-- One interesting extension of our
work is analysis of the timescale of formation of antiferromagnetic
order in optical lattices for interaction strength which are not too
large. In these cases, antiferromagnetic order relies on the BCS type
instability of the nested Fermi surface~\cite{Emery1976}. Therefore,
the quench dynamics is expected to have resemblances with the behavior
shown in Fig.~\ref{fig:BCS}.

{\it Cold atom experiments}-- Finally, we would like to comment on the
possibility of observing the described dynamics of texture growth in
realistic cold atom experiments. There are two issues here: (a) the
length scale corresponding to the most unstable mode and (b) the time
scale associated to the growth of unstable modes. The length scale
$1/q_{\mathrm{max}}$ must be sufficiently large so that patterns can
be resolved. Achieving large pattern sizes requires a quench to a
final coupling which is close to the critical value.  However, the
growth rate $\Delta_{\mathrm{max}}$ is small in this regime and the
textures develop over a long time. Therefore, one needs to find a
window of $U_f$, where the length scale is resolved in experiments
while the timescale is short enough to prevent considerable atom
losses. In recent experiments~\cite{MIT:Stoner:Expt}, the quench time
was $\sim 4\,\text{ms}$, which was followed by a hold time of upto $\sim 12
\, \text{ms}$; the imaging resolution was $\sim 10 k_F^{-1}$.

We note that the quench approximation is applicable for times shorter
than the corresponding (time dependent) inverse growth rate, i.e.
$\Delta_{\text{max}}[U(t)]^{-1} \sim t-t_c$, where $t_c$ is the time
at which the instability is crossed, $U(t_c)=U_c$. At longer times,
the evolution of the magnetization is dominated by the much slower
dynamics of the coalescence of the magnetic domains.  Assuming that
$U(t)$ in the MIT experiments is ramped up in time from $0$ to $U_f$
at a constant rate, the quench approximation is applicable to the MIT
experiment with final couplings of $U_f/U_c \lesssim 1.06$. For data
points where $U_f/U_c \gtrsim 1.06$, the quench dynamics stops at the
point $U(t)/U_c \sim 1.06$ which results in a pattern of magnetic
domains with typical size $\sim 2 k_F^{-1}$ that coarsens very slowly
over the remainder of the experment.  In the MIT data, the typical
step size in the final interaction strength was $\Delta k_F a_s \sim
0.3$, thus we would expect that the smallest $U_f$ that exceeded $U_c$
was $U_f\sim 1.06 \, U_c$. Therefore, we would not expect to see a
texture with a wavelength longer than $\sim 2 k_F^{-1}$.  This is
consistent with the fact that individual textures were not resolved in
the current set of experiments, as the MIT setup can only resolve
textures with wavelength of $\sim 10 k_F^{-1}$. However, our theory
indicates that domains of this size could be achieved if the quench is
stopped at $U_f/U_c \sim 1.004$, after a hold time of $\sim
14\,\text{ms}$, which could be possible with current experiments via a
finer tuning of the final interaction strength (although the
inhomogeneous density profile in the trap and three body
recombinations do make it difficult to observe).

{\it Acknowledgements} It is our pleasure to thank E. Altman, D. Huse,
M. Lukin, I. Mazin, S. Stringari, I. Carusotto, and specially W. Ketterle for useful
discussions. The authors acknowledge support of DARPA, CUA, and NSF Grant
No. DMR-07-05472 during this work.


\begin{thebibliography}{99}
\bibitem{init_footnote} Both of these conditions can be easily
  relaxed. However, the simple case already exhibits
  the important physical features.
\end{thebibliography}
\end{document}